\documentclass[10pt]{article}

\usepackage{amsmath}
\usepackage{array}
\usepackage{appendix}
\usepackage{graphicx}
\usepackage{amsfonts}
\usepackage{amssymb}
\usepackage{mathrsfs}
\usepackage{yfonts}
\usepackage{euscript}
\usepackage{upgreek}
\usepackage{slantsc}
\usepackage{calligra}
\usepackage[T1]{fontenc}
\usepackage{epsf}
\usepackage{latexsym}

\usepackage{tipa}

\def\be{\begin{equation}}
\def\ee{\end{equation}}
\def\beq{\begin{equation}}
\def\eeq{\end{equation}}
\def\bea{\begin{eqnarray}}
\def\eea{\end{eqnarray}}

\def\ni{\noindent}

\def\!{\hspace{-1.6667em}}

\def\mD{\mbox{D}}

\def\mF{\mbox{F}}
\def\mG{\mbox{G}}

\def\mJ{\mbox{J}}  
\def\mK{\mbox{K}}
\def\mL{\mbox{L}}
\def\mM{\mbox{M}}

\def\mP{\mbox{P}}
\def\mQ{\mbox{Q}}
\def\mR{\mbox{R}}
\def\mS{\mbox{S}}
 
\def\mU{\mbox{U}}

\def\mW{\mbox{W}}

\def\me{\mbox{e}}

\def\mh{\mbox{h}}

\def\mn{\mbox{n}}   
\def\mo{\mbox{o}}
\def\mp{\mbox{p}}

\def\ms{\mbox{s}}
\def\mt{\mbox{t}}

\def\ux{\underline{{x}}}

\def\bh{\underline{\underline{\mbox{h}}}  }            

\def\suF{\underline{\mbox{\scriptsize F}}}

\def\bh{\mbox{{\bf h}}}

\def\scH{\mbox{\scriptsize ${\cal H}$}}          

\def\scM{\mbox{\scriptsize ${\cal M}$}}          

\def\FrQ{\mbox{\Large $\mathfrak{q}$}}

\def\FrM{\mbox{\Large $\mathfrak{m}$}}                          
\def\sFG{\mbox{$\mathfrak{g}$}}

\def\FrG{\mbox{\Large $\mathfrak{g}$}}

\def\sa{\mbox{\scriptsize a}}
\def\sb{\mbox{\scriptsize b}}
\def\scc{\mbox{\scriptsize c}}

\def\se{\mbox{\scriptsize e}}

\def\sg{\mbox{\scriptsize g}} 
 
\def\si{\mbox{\scriptsize i}}

\def\sll{\mbox{\scriptsize l}}  
\def\sm{\mbox{\scriptsize m}}
\def\sn{\mbox{\scriptsize n}} 
\def\so{\mbox{\scriptsize o}}

\def\sr{\mbox{\scriptsize r}}
\def\st{\mbox{\scriptsize t}}

\def\sv{\mbox{\scriptsize v}}

\def\sB{\mbox{\scriptsize B}}

\def\sF{\mbox{\scriptsize F}}
\def\sG{\mbox{\scriptsize G}}

\def\sJ{\mbox{\scriptsize J}}

\def\sL{\mbox{\scriptsize L}}

\def\sP{\mbox{\scriptsize P}} 
 
\def\sR{\mbox{\scriptsize R}}
\def\sS{\mbox{\scriptsize S}}
\def\sT{\mbox{\scriptsize T}}

\def\sV{\mbox{\scriptsize V}}

\def\sfA{\mbox{\sffamily{\scriptsize A}}}
\def\sfB{\mbox{\sffamily{\scriptsize B}}}

\def\tn{\mbox{\tiny n}}

\def\K{Kucha\v{r} }

\def\pa{\partial}
\def\d{\textrm{d}}

\def\Circ{\mbox{\Large$\circ$}}               
\def\5Star{\mbox{\Large$\star$}}              
\def\cr{\mbox{\scriptsize{\bf $\mbox{ } \times \mbox{ }$}}}

\def\sumi3{\sum\mbox{}_{\mbox{}_{\mbox{\scriptsize $i$=1}}}^3}

\def\sumj3{\sum\mbox{}_{\mbox{}_{\mbox{\scriptsize $j$=1}}}^3}
\def\sumk3{\sum\mbox{}_{\mbox{}_{\mbox{\scriptsize $k$=1}}}^3}

\begin{document}

\begin{center}

{\LARGE\bf          PROBLEM OF TIME} 

\vspace{.1in}

{\LARGE\bf  IN SLIGHTLY INHOMOGENEOUS COSMOLOGY}

\vspace{.1in}

{\bf Edward Anderson}

\vspace{.1in}

{\em DAMTP Cambridge}, ea212 *at* cam.ac.uk

\end{center}

\begin{abstract}

The Problem of Time (PoT) is a multi-faceted conceptual incompatibility between various areas of Theoretical Physics.
Whilst usually stated as between GR and QM, in fact 8/9ths of it is already present at the classical level.  
Thus we adopt a `top--down' classical and then quantum approach.
I consider a local resolution to the Problem of Time that is Machian, which was previously realized for relational triangle and minisuperspace models.  
This resolution has three levels: classical, semiclassical and combined semiclassical--histories--records.
This article's specific model is a slightly inhomogeneous cosmology considered for now at the classical level.      
This is motivated by how the inhomogeneous fluctuations that underlie structure formation -- galaxies and CMB hotspots -- might have been seeded by quantum cosmological fluctuations, 
as magnified by some inflationary mechanism. 
In particular, I consider the perturbations about $\mathbb{S}^3$ case of this involving up to second order, 
which has a number of parallels with the Halliwell-Hawking model but has a number of conceptual differences and useful upgrades.
The article's main features are that the elimination part of the model's thin sandwich is straightforward, but the modewise split of the constraints fail to be first-class constraints.
Thus the elimination part only arises as an intermediate geometry between superspace and Riem.
The reduced geometries have surprising singularities influenced by the matter content of the universe, though the $N$-body problem anticipates these with its collinear singularities.
I also give a `basis set' of \K beables for this model arena.

\end{abstract}

\section{Introduction}\label{Intro}

The Problem of Time (PoT) \cite{Kuchar92I93, APoT, APoT2, FileR, APoT3} is a multi-faceted conceptual incompatibility between various areas of Theoretical Physics.
While this is usually stated as between GR and QM, in fact 8/9ths of it is already present at the classical level \cite{BI}.  
A `top down' classical and then quantum approach is adopted.  
Moreover, we arrive there by considering the conceptually and philosophically interesting case of background-independent physical theories \cite{BI}, which have the PoT as a consequence.  
The quantum version of the PoT is more severe, the almost-complete classical manifestation of the problem is expected to be a useful precursor as regards the 
form, and resolution, of the quantum version of the problem.  

\mbox{ }  

\ni We begin with the Temporal Relationalism facet of the PoT. 
Temporal Relationalism is the source of the well-known quantum-level Frozen Formalism Problem of the Wheeler--DeWitt equation \cite{Battelle, DeWitt}. 
This source is already classically present; it is the Leibnizian idea that there is no meaningful notion of time for the universe as a whole.  
Sec 2 then covers the following.  

\noindent 1) It explains that {\it Temporal Relationalism} can be mathematically implemented by manifest reparametrization invariance, manifest 
parametrization irrelevance, or geometrical actions that happen to be dual to the latter \cite{Lanczos, Magic, BSW, RWR, FileR, AM13}.

\noindent 2) Furthermore, Temporal Relationalism leads directly to the Hamiltonian constraint $\scH$.
Thus at the quantum level it leads to the Wheeler--DeWitt equation $\widehat{\scH}\Psi = 0$ (for $\Psi$ the wavefunction of the univcerse), 
which manifests the familiar Frozen Formalism Problem.  
[The Wheeler--DeWitt equation is to be contrasted with the time-dependent Schr\"{o}dinger equation and other time-dependent quantum wave equations.]  

\noindent 3) The primarily timeless situation is resolved at the classical level along the lines of Mach's time Principle -- time is to be abstracted from change $\d Q^{\sfA}$ 
(for configurations $Q^{\sfA}$ which form the configuration space $\FrQ$).  
This is in a manner that extends the concept of the astronomers' ephemeris time \cite{Clemence}.

Sec 3 then considers {\it Configurational Relationalism}, concerning the practical use of a group $\FrG$ of physically irrelevant transformations acting upon $\FrQ$.     
This is to be implemented by corrections to the changes (deparametrized velocities). 
One then extremizes the action with respect to the $\FrG$ auxiliaries (known as `Best Matching' \cite{BB82, FileR}, which is a type of reduction).  
Moreover, the output of this extremization features in the expression for $t^{\se\sm}$ for theories with nontrivial $\FrG$.

Minisuperspace does not manifest nontrivial Configurational Relationalism \cite{AMSS-1}, but relational particle mechanics (RPM) does \cite{BB82, FileR}, as follows.
The action for this can be expressed as 
\beq
S = 2\int\sqrt{WT}\d\lambda    \mbox{ } , \mbox{ } \mbox{ } 
W := E - V(\mbox{\boldmath$q$}) \mbox{ } , \mbox{ } \mbox{ } 
T := M_{iIjJ}\Circ_{A, B}q^{iI}\Circ_{A, B}q^{jJ}/2 \mbox{ } , \mbox{ } \mbox{ } 
\Circ_{\underline{A}, \underline{B}}\underline{q}^{I} := \dot{\underline{q}} - \dot{\underline{A}} - \dot{\underline{B}} \cr \underline{q}^I   \mbox{ } .  
\eeq
Here $W$ is the `potential factor', with constituent parts $V$ the potential energy and $E$ the total energy.
$T$ is the kinetic term, built out of the kinetic metric (alias in this case mass matrix) $M_{iIjJ} = \delta_{IJ}\delta_{ij}m_I$.
The expression given for $\Circ_{\underline{A}, \underline{B}}$ defines the best-matching derivative with respect to the translational auxiliary $\underline{A}$ 
                                                                                                                 and the rotational    auxiliary $\underline{B}$.
Underlining denotes spatial vector (also lower-case latin indices, whereas upper-case ones denote particle labels), 
and bold font denotes configuration space quantity (here possessing both of these types of indices).
Then varying with respect to $\underline{A}$ gives the zero total momentum         constraint $\underline{P} := \sum_I \underline{p}_I = 0$,
and          with respect to $\underline{B}$ gives the zero total angular momentum constraint $\underline{L} := \sum_I \underline{q}^I \cr \underline{p}_I  = 0$. 
Best Matching then involves solving the velocity formulation of these constraints for the extremal values of $\underline{A}$ and $\underline{B}$ themselves.  
In 2-$d$ these can be entirely solved for \cite{FileR}. 
[In 3-$d$ they can be solved for {\it locally}, meaning away from the physically bona fide collinear configurations 
for which, nonetheless, the configuration space geometry becomes singular.]

For full GR, $\FrQ$ = Riem($\Sigma$): the space of positive-definite 3-metrics on a fixed spatial topology $\Sigma$, and $\FrG$ = Diff($\Sigma$): the corresponding diffeomorphisms.  
Best Matching here then involves solving the so-called Thin Sandwich Problem \cite{WheelerGRT, BSW, TSC1TSC2FodorTh}: solving the linear momentum constraint for the GR shift 
with spatial metric $\mh_{ab}$ and its label-time velocity $\dot{\mh}_{ab}$ as data on a spatial hypersurface $\Sigma$.
The Thin Sandwich -- Fig 1.b) -- is the infinitesimal limit of the thick sandwich [Fig 1.a)], and features as a second facet of the PoT \cite{Kuchar92I93}.
Fig 1.c) recasts this in Best Matching form, which is more general over the set of theories [e.g. Fig 1.d) exhibits the corresponding Best Matching for the RPM triangle].
This Thin Sandwich Problem is in general a major unsolved problem \cite{TSC1TSC2FodorTh}; however, the current Article, demonstrates that it is surmountable for the practically relevant 
subcase of GR that is slightly inhomogeneous cosmology (SIC).    
Since slightly inhomogeneous quantum cosmology is a case of considerable interest (see below), this adds substantial value to quantum gravity schemes \cite{FileR, AHall, A13} 
that require Thin Sandwich resolution at an early stage.

Note that the above Lagrange multiplier implementation of Configurational Relationalism -- rooted in conventional Principles of Dynamics practise -- spoils Temporal Relationalism. 
This is resolved by a more careful choice of Configurational Relationalism's auxiliaries \cite{FEPI, FileR}, as per Sec 3.  
Take this as a first indication that attempted resolutions of individual PoT facets have a great tendency to interfere with each other \cite{Kuchar92I93, FileR, ABook}.
See \cite{TRiPoD} for a full modification of the Principles of Dynamics to be compatible with Temporal Relationalism.
That is the amount of work it takes to make even just a classical framework in which the other 
PoT facets can be formulated and strategically addressed without losing one's resolution of just one other facet.  
%
{            \begin{figure}[ht]
\centering
\includegraphics[width=0.65\textwidth]{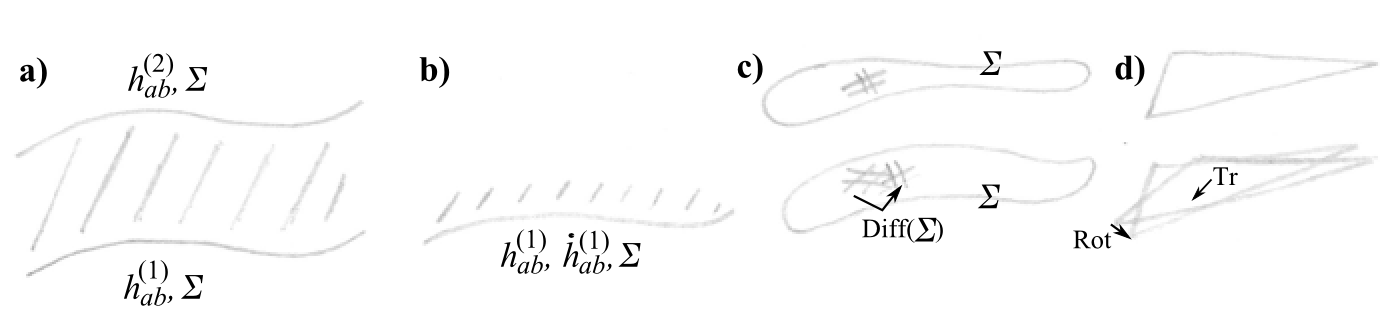}
\caption[Text der im Bilderverzeichnis auftaucht]{        \footnotesize{a) Thick sandwich and its thin-sandwich limit b).  
The data are as given and the problems to solve are for the spacetime in each shaded region.
c) Thin sandwich recast as a case of Best Matching. 
d) RPM triangle case of Best Matching.
For d) one considers two triangles, keeping one fixed whilst shuffling the other around using rotations Rot and translations Tr until maximal congruence is attained. 
For c) one considers two spatial geometries, keeping one fixed whilst shuffling the other around with spatial diffeomorphisms Diff($\Sigma$).} }
\label{CR-Facet} \end{figure}          }
 
This Article's model -- SIC -- combines temporal features of minisuperspace and RPM into one arena, 
and with the added benefit of being more cosmologically realistic than either model.
We set this model up in Secs 4 (configurations), 5 (the model's $\FrQ$ and $\FrG$) and 6 (the model's relational action).  
SIC is a perturbative treatment about some minisuperspace.  
In particular, this Article considers a second-order perturbation treatment about the spatially isotropic $\mathbb{S}^3$ model with a single minimally-coupled scalar field. 
This shares many features with Halliwell and Hawking's \cite{HallHaw} model; differences from this due to relationalism being followed in this Article are laid out in Secs 2 to 6.
I choose this model for the following reasons.

\noindent 1) Closed models are Machian, $\Sigma = \mathbb{S}^3$ is the simplest case, and the most conventional for closed-universe cosmologies. 

\noindent 2) One needs at least 2 degrees of freedom for a relational formulation of a physical system, 
for the relational minimum is that one physical quantity evolves in terms of another. 

\noindent 3) Cosmology conventionally makes use of scalar fields.  
The simplest case involves one scalar field.  
It is not hard to extend to n scalar fields as regards this Article's considerations.
A cosmological constant term is needed \cite{Rindler} to support the closed spatially-$\mathbb{S}^3$ homogeneous isotropic Friedmann--Lema\^{\i}tre--Robertson--Walker (FLRW) 
cosmology with scalar field matter in the case with matter effects are presumed small.
SIC matters foremost due to how the inhomogeneous fluctuations that underlie structure formation \cite{HallHaw, KieferBook, Kiefer87KS91KK11} 
-- galaxies and CMB hotspots -- might have been seeded by quantum cosmological fluctuations, as magnified by some inflationary mechanism \cite{Inf}. 
Moreover, this particularly practically relevant setting suffices to manifests all facets of the Problem of Time (PoT).
The constraints underlying many of these facets are provided in Sec 7 for the above model, 
with the indirect formulation of its Machian classical emergent time in Sec 8, and the sandwich equation and its aforementioned successful solution in Sec 9.

I next name the remaining PoT facets; see the Section ascribed to each for the facet's meaning plus the specific example of that facet in SIC.  
Constraint Closure \cite{Dirac} (Sec 10), Expression in terms of Beables \cite{ABeables} (Sec 11), Spacetime Relationalism (Sec 12), Foliation Independence (Sec 13) \cite{T73, HKT} and 
Spacetime Construction \cite{RWR, AM13} (Sec 14). 
Minisuperspace is trivial as regards the Configurational Relationalism, Constraint Closure, Foliation Dependence and Spacetime Construction facets of the PoT. 
On the other hand, RPM is trivial for the Spacetime Relationalism and Construction and Foliation Dependence facets.
SIC, however, is motivated by having all nine facets non-trivial [I demonstrate the eight of them that occur at the classical level in this Article.]
Thus this model arena serves as the successor of both of minisuperspace and RPM's qualitative insights into the PoT.  
Further successes with this model include: understanding the degrees of freedom count in terms of the constraint algebraic structure, 
and establishing a `basis set' of \K beables for this model. 
(In contrast with observables being quantities that {\it are observed}, beables are quantities that just {\it are}. 
This is more appropriate for whichever of Cosmology or closed-system QM \cite{Bell}.) 
[\K beables are quantities that commute with a theory's first-class linear constraints; this is as opposed to Dirac beables, which commute with all of a theory's first-class constraints, 
and thus in particular with GR's quadratic Hamiltonian constraint as well.]

Sec 15 conciudes with this work's frontiers, including an outline of these results' quantum counterparts.

\section{Temporal Relationalism}

Temporal Relationalism can be implemented by constructing actions as follows. 

\mbox{ } 

\noindent i) These actions are not to contain any extraneous times or extraneous time-like variables.

\mbox{ } 

\noindent [This is a necessary {\sl pre-requisite}, common to all three variants of the {\sl main} implementation discussed below.  
To be clear about the nature of the extraneous entities excluded, Newtonian time is an example of extraneous time 
                                                         and the  Arnowitt--Deser--Misner (ADM) lapse of GR $\upalpha$               is an example of extraneous time-like variable.]

\mbox{ } 														 
														 
\noindent ii) Time is not to be smuggled into such actions in the guise of a label either.

\mbox{ } 

\noindent Then a first formulation of ii) is for a label to be present but physically meaningless because it can be changed for any other (monotonically related) label 
without changing the physical content of the theory.   
I.e. the action in question is to be {\it manifestly reparametrization-invariant}. 

\mbox{ } 

\noindent Via its use of a label time $\lambda$, this is a relatively conventional presentation.  
Then for instance a primary notion of velocity can defined as the derivative with respect to $\lambda$:   

\noindent
\beq
\mbox{velocity} := \d \mbox{(configuration variable)}/\d (\mbox{label time}) \mbox{ i.e. } \mbox{ } \d Q^{\sfA}/\d\lambda \mbox{ } .
\eeq 
Next, one can straightforwardly build the kinetic term $T := ||\Circ \mbox{\boldmath$Q$}||_{\mbox{\scriptsize \boldmath$M$}}\mbox{}^2/2 := M_{\sfA\sfB} \Circ Q^{\sfA} \Circ Q^{\sfB}/2$. 
We assume for now that this takes the most physically standard form that is homogeneous quadratic in the velocities: `Jacobi-type' \cite{Lanczos}. 
I lift this assumption in \cite{FileR}.
The action is then 
\be
S := \int\d\lambda \, L   =   2\int\d\lambda\sqrt{{T}{W}} \mbox{ } ,    
\label{action}
\ee 
for `potential factor' $W = W(\mbox{\boldmath$Q$}) := E - V(\mbox{\boldmath$Q$})$ for mechanics and  := $R - 2\Lambda$ for GR.
[The latter is restricted to minisuperspace in the present Sec. 
See the next Sec for extension to full GR. 
$R = R(t)$ alone here is the Ricci 3-scalar and $\Lambda$ is the cosmological constant.]  

\mbox{ } 

\noindent A second implementation for ii) is that the action be {\it manifestly parametrization irrelevant}, i.e. making no use of $\lambda$.
Consequently, there is no primary notion of velocity, kinetic energy, Lagrangian, and more \cite{FileR, AM13}.     
Velocities have been supplanted at the primary level by differentials (`changes in configuration'):
\be
\d \mbox{(configuration variable)} \mbox{ i.e. } \d Q^{\sfA} \mbox{ } . 
\ee 
Then also kinetic energy has been supplanted by {\it kinetic arc element} 
\beq
\d s := ||\d \mbox{\boldmath$Q$}||_{\mbox{\scriptsize \boldmath$M$}} := \sqrt{M_{\sfA\sfB}(\mbox{\boldmath$Q$})\d Q^{\sfA}\d Q^{\sfB}} \mbox{ } ,   
\eeq
and Lagrangians by `Jacobian alias physical arc elements' 
\beq
\d J = \d s \sqrt{2W(\mbox{\boldmath$Q$})} \mbox{ } .  
\eeq
Note that the kinetic and Jacobi arc elements are related by just a conformal transformation.
Thus one has 
\beq
S := \int \d J \mbox{ } . 
\label{GeneralAction}
\eeq
I.e. viewed in terms of the physical $\d J$, one has a {\it geodesic principle}. 
So the problem of motion reduces to the problem of finding the geodesics associated with some geometry.
[In the present Article's case a Riemannian geometry, see \cite{FileR} for other examples.]    
On the other hand, in terms of the kinetic $\d s$, one has a Misner-type \cite{Magic} {\it parageodesic principle}  (i.e. geodesic modulo a conformal factor).  

\mbox{ } 

\noindent A third formulation of ii) follows the second formulation's steps too but is considered to be the construction of an action corresponding to a given geometry.
Thus no reference is ever made to the parameter that is, in any case, irrelevant. 
It is a further advance for background-independent physics to not name one's entities or techniques after physically-irrelevant properties.  
For the present Article's case, this is {\sl Jacobi's construction} of a mechanics from a given geometry that then plays the role of the corresponding configuration space geometry.

\mbox{ } 

\noindent  E.g. the minisuperspace form of this \cite{AMSS-1} for the example of relevance to this Article is the Misner-type action \cite{Magic}
\beq
S =     \mbox{$\frac{1}{2}$}\int \d s \, \sqrt{\overline{W}}                  \mbox{ } , \mbox{ }  \mbox{ }  
        \d s := \sqrt{\mbox{exp}(3\Omega)\{- \d\Omega^2 + \d\phi^2\}}                                                                    \mbox{ } , \mbox{ }  \mbox{ } 
\overline{W} : = \mbox{exp}(3\Omega)\{\mbox{exp}(-2\Omega) - V(\phi) - 2\Lambda\}   \mbox{ } .  
\eeq
Here $\Omega$ is the Misner variable, related to the usual scalefactor by $a = \mbox{exp}(\Omega)$, $\phi$ is the scalar field, and overline is the standard notation for densitization.  

\mbox{ } 

\ni At the classical level, 
Temporal Relationalism can be resolved along the lines of classical time being emergent at a secondary level via {\it Mach's Time Principle}: time is to be abstracted from change'.

Three distinct proposals to implement this then involve `any change' (Rovelli                                                     \cite{Rfqxi}), 
                                                                  `all change' (Barbour                                           \cite{Bfqxi}) and my 
										   	                      {\it sufficient totality of locally significant change (STLRC)} \cite{ARel2}.  
All three of these proposals have some sense in which they are `democratic' \cite{ARel2}: 
repsectively that any available change can be chosen, 
all changes are included, and 
all changes are given an opportunity to contribute.  
However, only the last two take into consideration how `some clocks are better than others' is an essential part of accurate timekeeping \cite{Bfqxi}.
Additionally, only the first and the third are operationally realizable.\footnote{For STLRC, whereas all change has {\sl the opportunity} to contribute,  
only those changes whose contributions lead to effects above the desired accuracy are actually kept in practise, by which it also manages to be both operationally well-defined 
and a provider of accurate timekeeping.
This is to be contrasted with the case of `all change', for which, since some of the universe's contents are but highly inaccurately known or completely unknown, 
one can not include `all change' in accurate or practical calculations.}
%
For these reasons, STLRC wins out.  
Then the time abstracted from this is a {\sl generalization} of the astronomers' {\it ephemeris time}  that emphasizes that such a procedure is in practise {\sl local}. 
Thus I term it a `GLET' (generalized local ephemeris time), and posit the specialization of the Machian emergent time resolution to `GLET is to be abstracted from STLRC'. 

\mbox{ } 

\noindent A specific implementation of a Machian emergent time is then as follows. 
It is a time that is distinguished by its simplification of the momentum--velocity relations and equations of motion using 
$\pa/\pa t^{\se\sm(\sJ)} := \sqrt{W/T}\pa/\pa\lambda = \sqrt{2W}\d /\d s$.  
This can be integrated up to give
\beq
t^{\se\sm(\sJ)} = \int \d\lambda \sqrt{T/W} = \int \d s \left/ \sqrt{2W} \right. \mbox{ } . 
\label{t-em-J}
\eeq
In the case of mechanics, this gives a recovery of Newton's time on a temporally-relational footing.
For the minisuperspace example of relevance to this Article, the emergent time takes the form 
\beq
t^{\se\sm(\sJ)} = \int  \sqrt{- \d\Omega^2 + \d\phi^2}\left/\sqrt{\mbox{exp}(-2\Omega) - V(\phi) - 2\Lambda} \right. \mbox{ } 
\label{plain-tem}
\eeq
and can be interpreted as a relational recovery of cosmic time.\footnote{(\ref{plain-tem}'s reality and monotonicity are assured by the Hamiltonian constraint, 
by which if the kinetic term in the numerator switches sign, then so does the potential factor in the denominator.
This also applies to the expression in (\ref{tem-SIC}).
These realizations of emergent time are motivated by the `GLET is to be abstracted from STLRC' realization of Mach's `time is to be abstracted from change' principle.}  

\section{Compatibility between Configurational and Temporal Relationalism}

\noindent Combining Temporal and Configurational Relationalism requires new auxiliaries.  
I.e. cyclic differentials in place of multiplier coordinates, with supporting free-end notion of space value variation \cite{FEPI, FileR}.\footnote{Cyclic differentials 
is the same useage as in the more common expression `cyclic velocities' in the Principles of Dynamics, the difference being that the former additionally evokes no (label) time.
I use $\d$ when acting on finite configuration variables ([$f(t)$ alone] and $\pa$ when acting on field variables [$F(t, x^i)$].
Finally, `notion of space' here means in particular `point' for particle models and `spatial hypersurface' for geometrodynamics and field theory.}
%
RPM and GR examples of doubly-relational actions are then, respectively, 
\beq
S = \sqrt{2} \int \d s\sqrt{E - V(\mbox{\boldmath$q$})}              \mbox{ } , \mbox{ } \mbox{ } 
\d s := ||\d_{\underline{A}, \underline{B}} \mbox{\boldmath$q$}||_{\mbox{\scriptsize\boldmath$M$}} := 
\sqrt{M_{iIjJ}\d_{\underline{A}, \underline{B}}q^{iI}\d_{\underline{A}, \underline{B}}q^{jJ}}                \mbox{ } , \mbox{ } \mbox{ } 
\d_{\underline{A}, \underline{B}} \underline{q}^{I} := \underline{q}^{I} - \d \underline{A} - \d \underline{B} \cr \underline{q}^{I}         \mbox{ } , \mbox{ } \mbox{ }
\eeq
\beq
S_{\sG\sR} = \int \int_{\Sigma}   \pa s_{\sG\sR} \sqrt{  \sqrt{\mh}\{\mR(\underline{x}; \bh] - 2\Lambda\}  }                   \mbox{ } , \mbox{ } \mbox{ }
\pa s_{\sG\sR} := ||\pa_{\underline{\sF}} \bh||_{\mbox{\scriptsize\boldmath${\cal M}$}} := \sqrt{  M^{ijkl}  \pa_{\underline{\sF}}\mh_{ij}  \pa_{\underline{\sF}}\mh_{kl}  }          \mbox{ } , \mbox{ } \mbox{ } 
\pa_{\underline{\sF}} \mh_{ij} := \pa \mh_{ij} - \pounds_{\pa \underline{\sF}} \mh_{ij}                   \mbox{ } . \mbox{ } \mbox{ }
\eeq
Here also $\underline{\mF}$ is a Diff($\Sigma$) auxiliary (`F for frame', such that $\dot{\underline{\mF}}$ is the usual ADM shift).  
Also,
\beq 
{\cal M}^{ijkl} := \sqrt{\mh}\{\mh^{ik}\mh^{jl} - \mh^{ij}\mh^{kl}\} \label{IDeWitt}
\eeq 
is the GR configuration space (alias inverse DeWitt) supermetric.
Straight fonts denote field quantities, and ( \, ; \, ] denotes a mix of function dependence (before the semicolon) and functional dependence (after it).

Then the RPM case of emergent Machian time -- Jacobi--Barbour--Bertotti (JBB) time \cite{BB82, FileR} -- is
\be
t^{\se\sm(\sJ\sB\sB)} = \mbox{\large E}_{\mbox{\scriptsize $\underline{A}, \underline{B}$  $\in$  Tr, Rot}}
\left(\int  ||\d_{\underline{A}, \underline{B}}\mbox{\boldmath$q$}||_{\mbox{\scriptsize\boldmath$M$}} \left/
                       \sqrt{ 2\{E - V(\mbox{\boldmath$q$})\}  }    \right.        \right) 
\label{TorreBruno}
\eeq
and the GR case is
\beq
\mt^{\se\sm(\sJ\sB\sB)}(\ux) = \mbox{\large E}_{\underline{\sF} \mbox{ }  \in  \mbox{ }  \mbox{\scriptsize Diff}(\Sigma)        }
\left(
\int \left. ||\pa_{\suF}\bh||_{\mbox{\boldmath\scriptsize${\cal M}$}} \right/   \sqrt{     \mR(\underline{x}; \bh] - 2\Lambda    }
\right)  \mbox{ } .
\label{GRemt2}
\eeq
Here $\mbox{\large E}_{g \in \sFG}$ denotes extremization over $\FrG$ of the corresponding relational action, subject to $\FrQ$, $\FrG$ and that action being suitably compatible 
\cite{ABook}.

\section{Configurations for SIC}

Parallelling \cite{HallHaw}'s treatment of the ADM split of GR for approximately homogeneous isotropic cosmologies, the 3-metric and scalar field are expanded as
\beq
\mh_{ij}(t, \underline{x}) = \mbox{exp}(2\Omega(t))\{S_{ij}(t) + \upepsilon_{ij}(t, \underline{x})\} 
\mbox{ } , \mbox{ } \mbox{ } 
\upphi(t, \underline{x}) = \sigma^{-1}\left\{\phi(t) + \upeta(t, \underline{x}) \right\} 
\mbox{ } . 
\label{Sepsi}
\eeq
Here, $S_{ij}$ is the standard hyperspherical $\mathbb{S}^3$ metric, and $\upepsilon_{ij}$ are inhomogeneous perturbations.
$\phi(t)$ is the homogeneous part of the scalar field, $\sigma := \sqrt{2/3\pi}/m_{\sP\sll}$ is a normalization factor, and $\upeta_{ij}$ are inhomogeneous perturbations.
The perturbation parts $\upepsilon_{ij}$ and $\upeta$ share an implicit small perturbation parameter $\epsilon$ as a factor.
The perturbations can furthermore be expanded as 
\beq
\upeta(t, \underline{x}) = \sum\mbox{}_{\mbox{}_{\mbox{\scriptsize $\sn, \sll, \sm$}}}f_{\sn\sll\sm} \, \mQ^{\sn}_{\sll\sm}(\underline{x})  \mbox{ } ,  
\label{f}
\eeq
\beq
\upepsilon_{ij}=\sum_{\sn,\sll,\sm} 
\big\{ 
\sqrt{\mbox{$\frac{2}{3}$}} a_{\sn\sll\sm}S_{ij}\mQ^{\sn}\mbox{}_{\sll\sm} + \sqrt{6}      b_{\sn\sll\sm}\{\mP_{ij}\}^{\sn}\mbox{}_{\sll\sm}         + 
\sqrt{2}\{c^{\so}_{\sn\sll\sm}\{\mS^{\so}_{ij}\}^{\sn}\mbox{}_{\sll\sm}    +         c^{\se}_{\sn\sll\sm}\{\mS^{\se}_{ij}\}^{\sn}\mbox{}_{\sll\sm}\} + 
      2 \{d^{\so}_{\sn\sll\sm}\{\mG^{\so}_{ij}\}^{\sn}\mbox{}_{\sll\sm}    +         d^{\se}_{\sn\sll\sm}\{\mG^{\se}_{ij}\}^{\sn}\mbox{}_{\sll\sm}\}  
\big\} .
\label{abcd}
\eeq
The superscripts `$\mo$' and `$\me$' stand for `odd' and `even'.  
n, l, m, $\mo$ and $\me$ mode labels are subsequently denoted by just a multi-index `n'.  
Following \cite{F35, GS78},            $\mQ_{\sn}(\underline{x})$                                      are the                                $\mathbb{S}^3$             scalar harmonics, 
                                       $\mS^{\so}_{\sn \, i}(\underline{x})$  and $\mS^{\se}_{\sn \, i}(\underline{x})$ are the transverse        $\mathbb{S}^3$             vector harmonics, 
						           and $\mG^{\so}_{\sn \, ij}(\underline{x})$ and $\mG^{\se}_{\sn \, ij}(\underline{x})$ are the transverse traceless $\mathbb{S}^3$ symmetric 2-tensor harmonics.

The     $\mS_{\sn \, ij}(\underline{x})$                    are then given by $\mS_{\sn \, ij} := \mD_j\mS_{\sn \, i} + \mD_i\mS_{\sn \, j}$  (for each of the $\mo$, $\me$ superscripts) 
and the $\mP_{\sn \, ij}(\underline{x})$  are traceless objects given by $\mP_{\sn \, ij} := \mD_j\mD_i\mQ_{\sn}/\{\mn^2 - 1\} + S_{ij}\mQ_{\sn}/3$. 
An important distinction to make is between the {\it plain perturbation scheme} of (\ref{Sepsi}) and the {\it modewise perturbation scheme} of (\ref{f}, \ref{abcd}).

In the latter, additionally, the relational formulation's differential of the frame auxiliary is expanded as
\beq
\pa \mF_i = \mbox{exp}(\Omega) \sum\mbox{}_{\mbox{}_{\mbox{\scriptsize $\sn, \sll, \sm$}}} 
\left\{
\d k_{\sn\sll\sm} \{\mP_i\}^{\sn}\mbox{}_{\sll\sm}/\sqrt{6} + 
\sqrt{2}  \{  \d j^{\so}_{\sn\sll\sm}  \{\mS^{\so}_i\}^{\sn}\mbox{}_{\sll\sm} + \d j^{\se}_{\sn\sll\sm}\{\mS^{\se}_i\}^{\sn}\mbox{}_{\sll\sm}  \}
\right\}
\eeq
for $\mP_{\sn \, i} := \mD_i\mQ_{\sn}/\{\mn^2 - 1\}$.  
This contains an overall $\epsilon$ factor due to the zeroth-order part now being zero.  
The relational formulation differs from \cite{HallHaw} not only in using this distinct formulation of auxiliary but also in not having a primary lapse to expand.
This is because the lapse is not held to have meaningful primary existence, so it is not to be an independent source of perturbations.  
Consequently the relational formulation has one family of coefficients less than Halliwell--Hawking (their $g_{\sn\sll\sm}$).

Note that {\it multipole expansion coefficients}  $a_{\sn}$, $b_{\sn}$, $c^{\so}_{\sn}$, $c^{\se}_{\sn}$, $d^{\so}_{\sn}$, 
$d^{\se}_{\sn}$, $f_{\sn}$, $\d j^{\so}_{\sn}$, $\d j^{\se}_{\sn}$, $\d k_{\sn}$ are functions of the coordinate time $t$ 
(which is also label time $\lambda$ for GR) alone.
I also use $x_{\sn}$ as a shorthand for the gravitational modes, $c_{\sn}$ and $d_{\sn}$ for each e-o pair of these, and $\d y_{\sn}$ for the auxiliaries considered together.  
The $x_{\sn}$, $f_{\sn}$ and $\d y_{\sn}$ are regarded as small quantities in the subsequent analysis, in particular with third-order quantities always neglected in this Article.  
%
{            \begin{figure}[ht]
\centering
\includegraphics[width=0.95\textwidth]{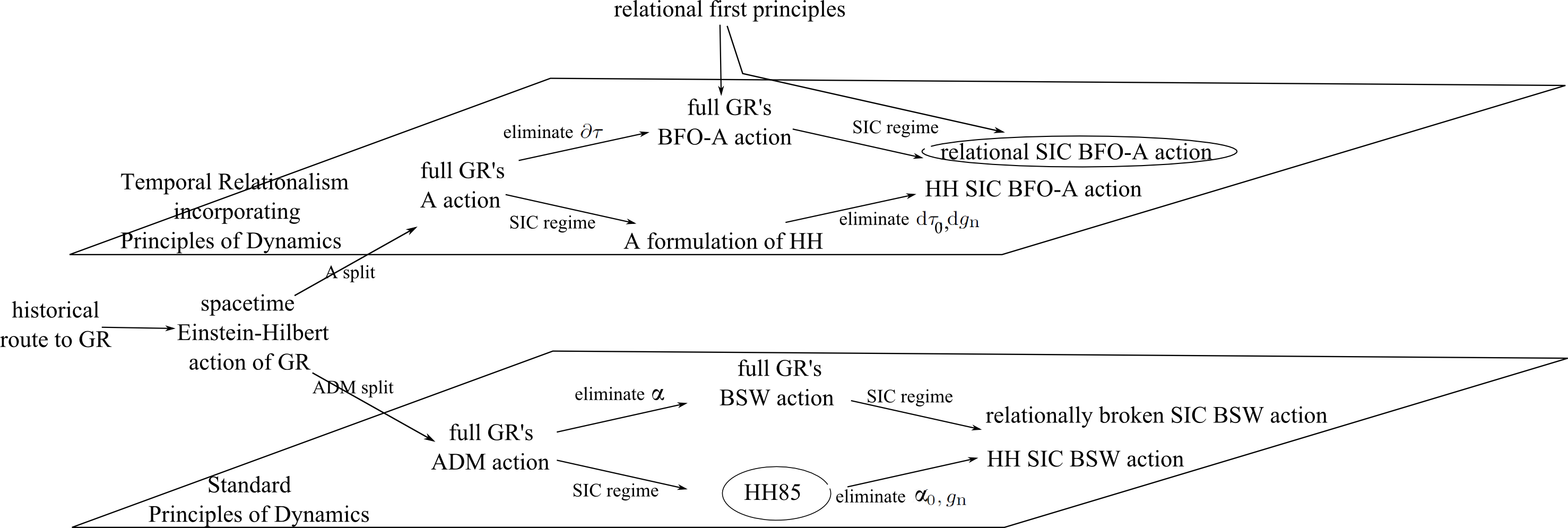}
\caption[Text der im Bilderverzeichnis auftaucht]{ \footnotesize{The differences between the relational and ADM--HH formulations of perturbative SIC. 
The differences between the upper and lower `floors' in the diagram involve the auxiliaries in use:    
Lagrange multipliers N, $N_0$, $g_{\sn}$ downstairs versus cyclic differentials $\pa \tau$, $\d \tau_0$ and $\d g_{\sn}$ upstairs.
HH stands for `Halliwell--Hawking', BSW for Baierlein--Sharp--Wheeler \cite{BSW}, BFO for Barbour--Foster--\'{o} Murchadha.  
The `A split' is the fully relational analogue of the ADM split, but using cyclic differential auxiliary variables $\pa \mF_i$ and $\pa\tau$ in place of ADM's shift and lapse 
auxiliary variables $\beta_i$ and $\alpha$ respectively.  
The bent arrow to the top encircled action is this Article's procedure, whereas the ADM--HH action is the second circled action.
Each of the upstairs and downstairs squares `do not commute', by which the two encircled actions are {\sl not} quite equivalent.  
They differ as regards how time is treated.
Nonetheless, they produce all of the same constraint equations in suitable Hamiltonian-type formulations. 
(For Temporal Relationalism compatibility \cite{AM13}, the relational approach requires a partial differential almost-Hamiltonian, 
i.e. an object that contains auxiliary partial derivative variables as well as configurations and momenta.)
This is with the exception of  the ADM--Halliwell--Hawking case containing a linear Hamiltonian constraint contribution from variation with respect to the perturbation of the lapse.  
}    }
\label{2-Levels} \end{figure}          }

\section{Outline of $\FrQ$ and $\FrG$ for SIC}

This arena's redundant configuration space is infinite-dimensional and spanned by $\Omega$, $\phi$, $x_{\sn}$ and $f_{\sn}$.
Counting this out, one has a 1-scalar FLRW minisuperspace's 2 global degrees of freedom and redundant GR's 6 degrees of freedom per value of the mode multi-index n plus the 
scalar's 1 degree of freedom per n. 
I.e. $2 + 7 \times \{\mbox{countable } \infty\}$ degrees of freedom.
Additionally, the $\d y_{\sn}$ contain the 3 (unphysical) degrees of freedom per n.
Thus the total redundant count is $2 + 10 \times \{\mbox{countable } \infty\}$ degrees of freedom.

For a pure-GR sector, Riem($\mathbb{S}^3$) is the $1 + 6 \times \{\mbox{countable } \infty\}$ dimensional space of scale variables and perturbations $x_{\sn}$.
The supermetric on this is readily computible from DeWitt's formula (\ref{IDeWitt}) via (\ref{Sepsi}) and (\ref{abcd}); the second-order contribution to this is as per 
Fig \ref{SIC-Metric}. 
Call this space equipped by this metric Riem$_{0,1,2}$($\mathbb{S}^3$)
Moreover, Fig 3 gives the full second order perturbed configuration space metric for GR with minimally-coupled scalar field. 
I.e. the $2 + 7 \times \{\mbox{countable } \infty\}$ dimensional space of scale variables, homogeneous scalar modes and perturbations $x_{\sn}, f_{\sn}$.
Fig 3 can display this and the pure GR sector at once because the scalar field is minimally coupled. 
Thus by the direct sum split for GR plus minimally-coupled matter (`mcm') \cite{Teitelboim}
\beq
{\cal M} = {\cal M}^{\sg\sr\sa\sv} \oplus {\cal M}^{\sm\scc\sm} \mbox{ } ,
\label{grav-oplus-mcm}
\eeq 
the SIC configuration space is $\FrQ_{0,1,2} = \mbox{Riem}_{0,1,2}(\mathbb{S}^3) \oplus \mbox{S}_{0,1,2}(\mathbb{S}^3)$ applies, for S the scalar field configuration space.  
Moreover, metric variables enter the scalar field sector but scalar field variables do not enter the gravitational sector.  
%
{\begin{figure}[ht]
\centering
\includegraphics[width=0.65\textwidth]{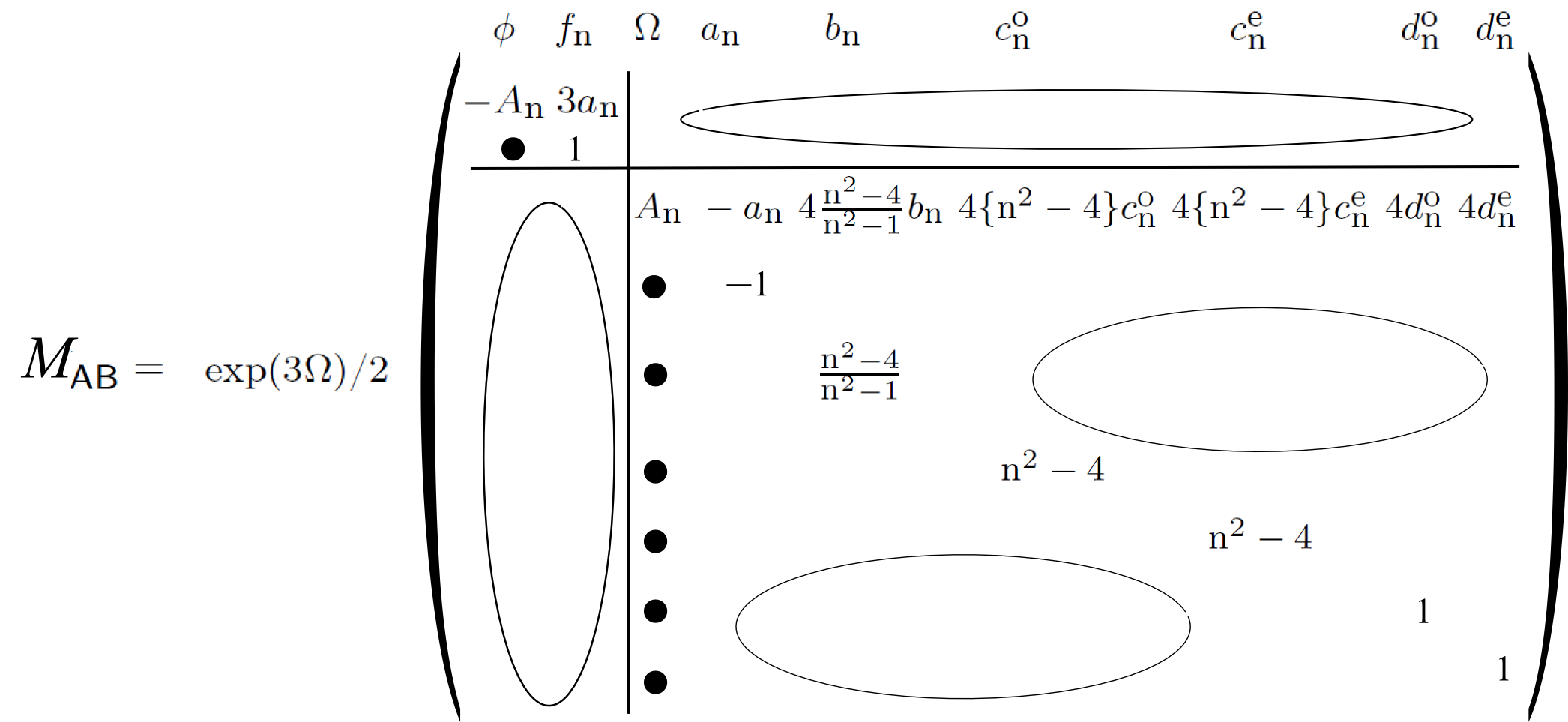}
\caption[Text der im Bilderverzeichnis auftaucht]{\footnotesize{SIC's configuration space metric.  
The heavy dot denotes `same as the transposed element' since metrics are symmetric.  
Note the matter--gravitation direct sum structure.
Note also the further `arrowhead-shaped' sparseness in the gravitational block, 
whose significance is that the gravitational modes only couple to this order to the homogeneous scale rather than also amongst themselves. }} 
\label{SIC-Metric}\end{figure}} 

\noindent  Note that there is no spatial dependence once one enters modewise equations.

\mbox{ } 

\noindent $\FrG =$ Diff($\mathbb{S}^3$) start to have effect at first order; I denote the corresponding space of $y_{\sn}$'s by Diff$_1$($\mathbb{S}^3$).

\section{Relational action for SIC}

At the modewise level, the relational action for this is 
\beq
S_{\sr\se\sll\sa\st\si\so\sn\sa\sll} = \sqrt{2} \int \d s_{0,1,2} \sqrt{     \overline{W}_{0,2}    }  \mbox{ } . 
\eeq
Here 
$\d s_{0,1,2}\mbox{}^2 = 
 {\sum\mbox{}_{\mbox{}_{\mbox{\scriptsize n}}}[\d_{\underline{\sF}} f_{\sn}, \d \phi, \d \Omega, \d_{\underline{\sF}} x_{\sn}]
            [{\cal M}_{\upphi}\oplus {\cal M}_{\sg\sr\sa\sv}]
			[\d_{\underline{\sF}} f_{\sn}, \d \phi, \d \Omega, \d_{\underline{\sF}} x_{\sn}]^{\st\sr}}$ 
This turns out to be the sum of `0' and `2' parts (below). 
The non-auxiliary portion of the `2' part can be read off Fig \ref{SIC-Metric},   
whereas the auxiliary terms match \cite{HallHaw}'s Lagrangian's under the correspondence $j_{\sn}/N_0 \rightarrow \d j_{\sn}$,  $k_{\sn}/N_0 \rightarrow \d k_{\sn}$.
In full,  
\beq
\d s_0\mbox{}^2 := {\mbox{exp}(3\Omega)\{- \d\Omega^2 + \d\phi^2\}}                                                                    \mbox{ } , \mbox{ }  \mbox{ } 
\eeq
$$
\d s_{2}^{\sn\,2} =  \frac{\mbox{exp}(3\Omega)}{2} 
\left\{
-\d{a}_{\sn}^2 + \frac{\mn^2 - 4}{\mn^2 - 1}\d{b}_{\sn}^2 + \{\mn^2 - 4\}\d{c}_{\sn}^2 + \d{d}^2_{\sn} + \d{f}_{\sn}^2 + 6a_{\sn}\d{f}_{\sn}\d{\phi} + \frac{2}{3}\d A_{\sn}\d{\Omega} + A_{\sn}\{\d{\Omega}^2 - \d{\phi}^2\}  
\right\}
$$
\beq
- \mbox{exp}(2\Omega)
\left\{
\{\mn^2 - 4\} \d{c}_{\sn} \d j_{\sn} + 
\left\{
\d{a}_{\sn} + \frac{\mn^2 - 4}{\mn^2 - 1}\d{b}_{\sn} + 3f_{\sn}  \d{\phi}
\right\}
\frac{\d k_{\sn}}{3}     
\right\} 
+ \frac{\mbox{exp}(\Omega)}{2}
\left\{
\{\mn^2 - 4\}\d j_{\sn}^2 - \frac{\d k_{\sn}^2}{3\{\mn^2 - 1\}} 
\right\} 
\mbox{ } .
\label{ds-012}
\eeq
This is expressed in terms of the useful combination (c.f. \cite{Wada})
\beq
A_{\sn} := -\frac{3}{2}
\left\{
a_{\sn}^2 - 4
\left\{ 
\frac{\mn^2 - 4}{\mn^2 - 1}b_{\sn}^2 + \{\mn^2 - 4\}c_{\sn}^2 + d_{\sn}^2
\right\}
\right\}  \mbox{ } .  
\label{An}
\eeq
This can be interpreted as follows.

\ni I) It is a gravitational sector configuration space volume correction term (from expanding the determinant).

\ni II) It is also the sole coupling to the FLRW minisuperspace degrees of freedom $\Omega$ and $\phi$.  

\ni N.B. it is not here being used as a coordinate.

One can then decompose (\ref{ds-012}) into whichever of S, V, T pieces or zeroth-, first- and second-order pieces. 
These pieces are readily visible in the quadratic form due to being labelled by $a_{\sn}, b_{\sn}, f_{\sn}$ factors, $c_{\sn}$ factors and $d_{\sn}$ factors on the one hand, 
and by how many powers of $\d x_{\sn}$, $\d y_{\sn}$ each term contains.

Also the densitized $\overline{W}_{0,2} = \overline{W}_0 + \sum\mbox{}_{\mbox{}_{\mbox{\scriptsize n}}} \overline{W}_2^{\sn}$ for $\overline{W}_0$ given by 
\beq
\overline{W}_0 : = \mbox{exp}(3\Omega)\{\mbox{exp}(\Omega) - V(\phi) - 2\Lambda\}                                                                       \mbox{ } . 
\label{MSS-Action}
\eeq
and 
$$
\overline{W}^{\sn}_2      = \frac{\mbox{exp}(\Omega)}{2}
\left\{
\frac{1}{3}
\left\{
\mn^2 - \frac{5}{2}
\right\}
a_{\sn}^2 + \frac{\{\mn^2 - 7\}}{3}\frac{\{\mn^2 - 4\}}{\mn^2 - 1}b_{\sn}^2 + \frac{2}{3}\{\mn^2 - 4\}a_{\sn}b_{\sn} - 2\{\mn^2 - 4\}c_{\sn}^2 - \{\mn^2 + 1\}d_{\sn}^2  \right\} 
$$
\beq
     +\frac{\mbox{exp}(3\Omega)}{2}
\left\{
- m^2\{f_{\sn}^2 + 6a_{\sn}f_{\sn}\phi\} - \mbox{exp}(-2\Omega)\{\mn^2 - 1\}f_{\sn}^2
- \{m^2\phi^2 + 2\Lambda\}A_{\sn}
\right\} \mbox{ } .
\label{W-n}
\eeq
[The first line is the nth mode's second-order contribution to the densitized Ricci scalar $\bar{R}_2^{\sn}$, 
whereas the second line comprises the matter potential and cosmological constant contributions.]  

\mbox{ }

\noindent Out of usefulness of this in subsequent cosmological modelling, relative to \cite{HallHaw} itself, I have i) added a cosmological constant term. 
ii) This Article's model and PoT workings also immediately extend to the case with $p$ minimally-coupled scalar fields.

\section{Constraints for relational SIC}

At the modewise level, the classical Hamiltonian constraint gives 
\beq
\scH := \mbox{$\frac{\mbox{\scriptsize exp}(-3\Omega)}{2}$}\big\{-\pi_{\Omega}^2 + \pi_{\phi}^2 + \mbox{exp}(6\Omega)\{V(\phi) + 2\Lambda - \mbox{exp}(-2\Omega)\big\} = 0 \mbox{ } . 
\label{MSS-H}
\eeq 
at zeroth order, and, at second order,
\beq
{\scH}_2 = \sum\mbox{}_{\mbox{}_{\mbox{}_{\mbox{}_{\mbox{\scriptsize n}}}}} \left\{ \, \mbox{}^{\sS}{\scH}_2^{\sn} +  \mbox{}^{\sV}{\scH}_2^{\sn} + \mbox{}^{\sT}{\scH}_2^{\sn} \, \right\} 
\mbox{ } \mbox{ for } 
\eeq
$$
^{\sS}{\scH}_2^{\sn} =  \mbox{$\frac{\mbox{\scriptsize exp}(-3\Omega)}{2}$}
\left\{
\left\{
\mbox{$\frac{1}{2}$}a_{\sn}^2 + 10\mbox{$\frac{\sn^2-4}{\sn^2-1}$}b_{\sn}^2
\right\}
\pi_{\Omega}^2+
\left\{
\mbox{$\frac{15}{2}$}a_{\sn}^2+6\mbox{$\frac{\sn^2-4}{\sn^2-1}$}b_{\sn}^2
\right\}
\pi_{\phi}^2-\pi_{a_{\tn}}^2+\mbox{$\frac{\sn^2-1}{\sn^2-4}$}\pi_{b_{\tn}}^2 + \pi_{f_{\tn}}^2 + 2a_{\sn}\pi_{a_{\tn}}\pi_{\Omega} + 8 b_{\sn}\pi_{b_{\tn}}\pi_{\Omega}- 6 a_{\sn}\pi_{f_{\tn}}\pi_{\phi} 
\right\}
$$
$$
-\mbox{$\frac{\mbox{\scriptsize exp}(\Omega)}{2}$}
\left\{
\mbox{$\frac{1}{3}$}
\left\{
\mn^2 - \mbox{$\frac{5}{2}$}
\right\}
a_{\sn}^2 +  \mbox{$\frac{\{\sn^2 - 7\}}{3} \frac{\{\sn^2 - 4\}}{\sn^2 - 1}$}b_{\sn}^2 + \mbox{$\frac{2}{3}$}\{\mn^2 - 4\}a_{\sn}b_{\sn} - \{\mn^2 - 1\}f_{\sn}^2
\right\}
$$
\beq
+\mbox{$\frac{\mbox{\scriptsize exp}(3\Omega)}{2}$}
\left\{
m^2\{f_{\sn}^2 + 6a_{\sn}f_{\sn}\phi\} + \{m^2\phi^2 + 2\Lambda\} 
\left\{
\mbox{$\frac{3}{2}$}a_{\sn}^2 - 6\mbox{$\frac{\sn^2 - 4}{\sn^2 - 1}$}b_{\sn}^2
\right\}
\right\}
\mbox{ } ,  
\eeq
\beq
^{\sV}{\scH}_2^{\sn} = \mbox{$\frac{\mbox{\scriptsize exp}(-3\Omega)}{2}$}
\left\{
\{\mn^2 - 4\} c_{\sn}^2 \{10\pi_{\Omega}^2 + 6\pi_{\phi}^2\} + \mbox{$\frac{\pi_{c_{\tn}}^2}{\sn^2 - 4}$} + 8c_{\sn}\pi_{c_{\tn}}\pi_{\Omega} 
\right\}
+ \{\mn^2 - 4\}c_{\sn}^2\{\mbox{exp}(\Omega) - 3\mbox{exp}(3\Omega)\{m^2\phi^2 + 2\Lambda\}\}
\mbox{ } , 
\eeq
\beq
^{\sT}{\scH}_2^{\sn} =  \mbox{$\frac{\mbox{\scriptsize exp}(-3\Omega)}{2}$}
\left\{ 
d_{\sn}^2\{10\pi_{\Omega}^2 + 6\pi_{\phi}^2\}  + \pi_{d_{\tn}}^2 + 8d_{\sn}\pi_{d_{\tn}}\pi_{\Omega}
\right\}
+ d_{\sn}^2
\left\{
\mbox{$\frac{ \sn^2 + 1 }{2}$}\mbox{exp}(\Omega) - 3\mbox{exp}(3\Omega)\{m^2\phi^2 + 2\Lambda \}
\right\}                                                                                \mbox{ } .
\eeq
\noindent Also $\scM_{1i} = [ \mbox{}^{\sS}\scM_{1}^{\sn}, \mbox{ }^{\sV}\scM_1^{\sn}]$ 
is the vector of constraints corresponding to the $[\d k_{\sn}, \d j_{\sn}^{\so}, \d j_{\sn}^{\se}]$ of auxiliaries.
The momentum constraint vanishes at zeroth order, and has S and V parts to first order:
\beq
^{\sS}{\scM}_1^{\sn} = \mbox{$\frac{\mbox{\scriptsize exp}(-3\Omega)}{3}$}
\left\{
- \pi_{a_{\tn}} + \pi_{b_{\tn}} + 
\left\{
a_{\sn} + 4\mbox{$\frac{\sn^2 - 4}{\sn^2 - 1}$}b_{\sn}
\right\}
\pi_{\Omega} + 3f_{\sn}\pi_{\phi}
\right\} \mbox{ } ,  
\label{M-S}
\eeq
\beq
^{\sV}{\scM}_1^{\sn} = \mbox{exp}(-\Omega)\{\pi_{c_{\tn}} + 4\{\mn^2 - 4\}c_{\sn}\pi_{\Omega}\} \mbox{ } .  
\label{M-V}
\eeq
\noindent Note 1) in Hamiltonian variables, the second-order Hamiltonian constraint pieces and the first-order momentum constraint pieces coincide for the relational and 
ADM--Halliwell--Hawking approaches.

\noindent Note 2) On the other hand,  a plain perturbation treatment produces $\underline{x}$-dependent constraints.

\section{Machian classical emergent time for SIC}

In the modewise approximation, the emergent JBB time is now 
\beq
t^{\se\sm(\sJ\sB\sB)} = \mbox{\large E}_{\d k_{\sn}, \d j_{\sn} \in \mbox{\scriptsize Diff$_1(\mathbb{S}^3)$}} 
\left(
\int\sqrt{\d s_0\mbox{}^2 +  \sum\mbox{}_{\mbox{}_{\mbox{}_{\mbox{}_{\mbox{\scriptsize n}}}}}\d_{j_{\sn}, k_{\sn}} s_{\sn}\mbox{}^2}
\left/
\sqrt{2\left\{W_0  + \sum\mbox{}_{\mbox{}_{\mbox{}_{\mbox{}_{\mbox{\scriptsize n}}}}} W_{\sn}\right\}}
\right.
\right)  
\mbox{ } ;  
\label{tem-SIC}
\eeq
{\large E} involves all n rather than just a particular n.

\section{SIC's thin sandwich}

Best Matching is the $Q^{\sfA}, \d Q^{\sfA}$ variables level solution of the constraints that are linear in the momenta. 
This generalizes the Thin Sandwich Problem to a wider variety of linear first-class constraints. 
Using $t^{\se\sm}$ to resolve the PoT specifically further motivates study of the thin sandwich formulation.  
The SIC case of this pushes one into a situation in which the reduction involves second-class constraints. 
Here na\"{\i}ve elimination does not suffice. 
One needs a further procedure. 
Gauge-fixing, second class constraint removing or reduction are candidates for such procedures. 
See \cite{AHH-2} for the third of these, with a study of the geometries of Riem, Superspace and spaces intermediate between these that occur in some of these routes.

All the momentum constraint components (\ref{M-S}, \ref{M-V}) are manifestly algebraic, and, being linear, manifestly solvable.
To address the Thin Sandwich Problem, these need to be recast in the formulation following from the Jacobi action principle. 
This is a configuration--change formulation: in terms of $Q^{\sfA}$ and $\d Q^{\sfA}$.
To pass from a constraint in configuration--momentum $Q^{\sfA}$ and $P_{\sfA}$ variables to its form in configuration--change variables, one uses the momentum--change relations.
These are analogous to the momentum--velocity relations used in passing from the Hamiltonian form of a constraint to its Lagrangian form (now in terms of $Q^{\sfA}$ and $\dot{Q}^{\sfA}$.

Then in the particular case of slightly inhomogenous cosmology, applying momentum--change relations to the S and V parts of the GR momentum constraint (\ref{M-S}, \ref{M-V}) 
gives the SIC thin sandwich equations
\be
\d a_{\sn} + \frac{\mn^2 - 4}{\mn^2 - 1}\d b_{\sn} + \mbox{exp}(-\Omega)\frac{\d k_{\sn}}{\mn^2 - 1} + 3f_{\sn}\d\phi = 0 \mbox{ } , 
\label{TSE-S}
\ee
\be
\d c_{\sn} - \mbox{exp}(-\Omega) \d j_{\sn}  = 0 \mbox{ } .  
\label{TSE-V}
\ee
These are to be interpreted as to be solved for the first-order auxiliary variables $\d j_{\sn}$ and $\d k_{\sn}$. 
The actual solving is in this case immediate.  
Just note that the above two equations are i) decoupled and ii) individually well-determined.  
[There are even and odd $\d j_{\sn}$ but also even and odd $\d c_{\sn}$, whilst everything in the equation for the single $\d k_{\sn}$ comes in a single copy.]
iii) They are algebraically trivial as regards solving by making whichever object therein the subject.
The solutions are
\beq
\d k_{\sn} = - \mbox{exp}(\Omega)  \big\{  \{  \d a_{\sn} + 3 f_{\sn} \d\phi  \}\{  \mn^2 - 1  \} + \{  \mn^2 - 4  \} \d b_{\sn}   \big\} \mbox{ } ,  
\label{TS-Soln-S}
\eeq
\beq
\d j_{\sn} = \mbox{exp}(\Omega)\d c_{\sn} \mbox{ } .
\label{TS-Soln-V}
\eeq
\noindent The bulk of the Thin Sandwich or Best Matching approach's work is, however, as follows. 

\ni 1) Substituting this back into the action to obtain a reduced action.

\ni 2) Obtain an explicit expression for the emergent time.
Via the $\d c_{\sn}$ and $\d j_{\sn}$ terms forming a square that cancels out by the constraint and the elimination of the $\d k_{\sn}$ producing another square in $\d a_{\sn} + \d b_{\sn}$ 
that is incorporated via the new `scalar sum' coordinate 
\beq
s_{\sn} := \sqrt{\{\mn^2 - 4\}/3}\{a_{\sn} + b_{\sn}\} \mbox{ } . 
\eeq
In this manner, the reduced configuration space's line element is [for $A_{\sn}$ now a coordinate]
\beq
\d s_{\sb\sm}^{\sn}\mbox{}^2 = \mbox{$\frac{\mbox{\scriptsize exp}(3\Omega)}{2}$}
\left\{  
\d s_{\sn}^2 + \d f_{\sn}^2 + \d d_{\sn}^2 +  
\left\{ 
\left\{
3 \d a_{\sn}  + \sqrt{3\{\mn^2 - 4\}} \d s_{\sn}
\right\}                                     
f_{\sn} +  6    a_{\sn} \d f_{\sn}
\right\}  \,
\d\phi 
+ \mbox{$\frac{2}{3}$}\d A_{\sn} \d\Omega - A_{\sn}\{ - \d\Omega^2 + \d\phi^2 \} 
\right\} \mbox{ } .  
\label{SIC-dsbm}
\eeq
Note that this is not superspace.  
This non-coincidence is tied to $^{\sS}\scM$ and $^{\sV}\scM$ not being first-class constraints, as further exposited in the next Sec.  
For now, I note that the block by block configuration space degrees of freedom count is 12 -- 2 -- 1 $\times$ 2 = 8.
The corresponding full configuration space degrees of freedom count is 2 + 10 N -- \{2 + 1 $\times$ 2\}N = 2 + 6N.  
This does not match superspace's degrees of freedom count, which involves taking out a $Diff(\Sigma)$ amount of variables: 3 $\times$ 2 per space point.  
It is then interesting to consider what geometry these intermediate-redundacy configuration spaces; see \cite{AHH-2} for a start on this.  
%
%

\section{Constraint closure for slightly inhomogenous cosmology}\label{SIC-CA}

{\bf Functional Evolution Problem} is Isham and Kucha\v{r}'s \cite{Kuchar92I93} name for the field theoric quantum-level case.     
This concerns whether the Hamiltonian and momentum constraints are all the constraints that one needs at the quantum level by virtue of constraints closing and of anomalies not arising.
The name `Constraint Closure Problem', however, is more widely applicable, both to finite models and at the classical level. 

\mbox{ } 

\noindent In the case of full GR at the classical level, this is resolved by the Dirac algebroid of constraints 
\be
\mbox{\bf \{} 
(    \scM_i    |    \pa \mL^i    ) 
\mbox{\bf ,} \, 
(    \scM_j    |    \pa \mM^j    ) 
\mbox{\bf \}} =  
(    \scM_i    | \, [ \pa \mL, \pa \mM ]^i )  \mbox{ } ,
\label{Mom,Mom}
\ee
\be
\mbox{\bf \{} 
(    \scH    |    \pa \mJ    ) 
\mbox{\bf ,} \, 
(    \scM_i  |    \pa \mL^i    ) 
\mbox{\bf \}} = 
(    \pounds_{\pa \underline{\sL}} \scH    |    \pa\mJ    )  \mbox{ } , 
\label{Ham,Mom}
\ee
\be 
\mbox{\bf \{} 
(    \scH    |    \pa \mJ    ) 
\mbox{\bf ,} \,
(    \scH    |    \pa \mK    )
\mbox{\bf \}}  = 
(    \scM_i \mh^{ij}   |    \pa \mJ \, \overleftrightarrow{\pa}_j \pa \mK    )  \mbox{ } . 
\label{Ham,Ham}
\ee
This is a standard algebraic result, modulo the form of the smearing functions used.
Smeared formulations are used for well-definedness in the sense of distributions.
Usually plain smearing functions rather than partial differentials are used, however the former spoil Temporal Relationalism compatibility \cite{AM13}.  
More specifically, I use $\pa \mJ$, $\pa \mK$ as smearing functions for $\scH$ and $\pa \mL^i$, $\pa \mM^i$ as smearing functions for $\scM_i$.   
I also use the notation $X \overleftrightarrow{\pa}^i Y := \{ \pa^i Y \} X - Y \pa^i X$ familiar from QFT.  

\mbox{ } 

\noindent Let us consider the case of plain perturbation theory (as opposed to applying modewise splits as well).
We work to second order overall, indicated by the 2-subscript notation.  
This means e.g. considering $\scM_{i0} + \scM_{i1}$ smeared by $\pa \mL^i_{0} + \pa \mL^i_{0}$ in the first bracket given.
Second-order inputs are here unecessary since $\scM_{i0} = 0$, by which the first factor is at least first-order, 
by which its smearing can at most be first-order if the computation is to be second-order overall.
Continuing this reasoning, the second-order brackets which arise are as follows.  

\be
\mbox{\bf \{} 
(    \scM_i    |    \pa \mL^i    ) 
\mbox{\bf ,} \, 
(    \scM_j    |    \pa \mM^j    ) 
\mbox{\bf \}}_2 =  
(    \scM_{i1}    | \, [ \pa \mL_1, \d \mM_0 ]^i + [ \d \mL_0, \pa \mM_1 ]^i)  \mbox{ } ,
\label{Mom,Mom-2}
\ee
\be
\mbox{\bf \{} 
(    \scH    |    \pa \mJ    ) 
\mbox{\bf ,} \, 
(    \scM_i  |    \pa \mL^i    ) 
\mbox{\bf \}}_2 = 
(    \pounds_{\pa \underline{\sL}_1} \scH_1    |    \d \mJ_0    )  \mbox{ } , 
\label{Ham,Mom-2}
\ee
\be 
\mbox{\bf \{} 
(    \scH    |    \d \mJ    ) 
\mbox{\bf ,} \,
(    \scH    |    \d \mK    )
\mbox{\bf \}}_2  = 
(    \scM_i   |   S^{ij} \{ \d\mK_0 \pa_j \, \pa\mJ_1  - \d\mJ_0 \pa_j \, \pa\mK_1 \}    )  \mbox{ } . 
\label{Ham,Ham-2}
\ee
\noindent \mbox{ } \mbox{ } SIC's modewise constraints require no smearing because they are finite block by block.  
By straightforward computation S, V, T cross-brackets are straightforwardly zero or weakly zero to second order.
%
%
Brackets between blocks of different $n$ are also straightforwardly zero.
All this is saying is that in calculations which are at most second-order overall, 
the modewise split into different values of $n$ and the S--V--T split are preserved under the brackets operation.  
Due to this, to at most second order overall, each S, V, T piece for each value of $\mn$ can be treated as a separate finite system in its own right, 
i.e. without reference to the other such systems.

Moreover, with each block being finite, self-brackets are all zero therein.
This gives immediately furthermore that $\mbox{\bf \{} \scH  \mbox{\bf ,} \, \scH \mbox{\bf \}}_2 = 0$ for each S, V, T piece and each $\mn$, 
and $\mbox{\bf \{} \scM  \mbox{\bf ,} \,  \scM \mbox{\bf \}}_2 = 0$ likewise.  
This leaves just two cross-brackets to evaluate: $\mbox{\bf \{} \mbox{}^{\sS}\scH \mbox{\bf ,} \, \mbox{}^{\sS}\scM \mbox{\bf \}}_2$ 
and $\mbox{\bf \{} \mbox{}^{\sV}\scH  \mbox{\bf ,} \, \mbox{}^{\sV}\scM \mbox{\bf \}}_2$.
[There is no $\mbox{}^{\sT}\scM$ to the order considered, so evaluating $\mbox{\bf \{} \mbox{}^{\sT}\scH \mbox{\bf ,} \, \mbox{}^{\sT}\scM \mbox{\bf \}}_2$ is a moot point.] 
Moreover, straightforward evaluation of these two cross-brackets produces nontrivial right hand side terms, in a manner which implies that 
$^{\sS} \scM$ and $^{\sS} \scH$ are not first-class constraints with respect to each other, and neither are $^{\sV} \scM$ and $^{\sV} \scH$.

See \cite{AHH-2} for a further phase space analysis of the corresponding breakdown of first-classness in the simpler vacuum case.
In that case, one is short by a lesser amount of degrees of freedom, 
and it turns out that this is reflected by the $\mbox{}^{\sV} \scH$ piece of $\scH$ ceasing to play the role of a first-class constraint.
This is, rather, an equation specifying the form of the corresponding vectorial part of the $\pa \mL^i$ auxiliary.  
Such `specifier equations' are a possibility which was indeed anticipated in Dirac's algorithm \cite{Dirac} for treating constrained systems.

In the case with minimally-coupled scalar field, one is short by one further degree of freedom, and additionally the fortunate decoupling by which $\mbox{}^{\sV} \scH$ 
becomes isolated as a specifier equation for the corresponding vectorial part of the $\pa \mF^i$ auxiliary does not occur.  
Due to this, full Principles of Dynamics and geometrical understandings of how the S, V, T split constraints for the minimally-coupled scalar field model 
cease to all be first-class has not yet been worked out for this model. 
None the less, \cite{AHH-2}'s detailed workings for the simpler vaccum case already suffice as confirmation that S, V, T splitting indeed need not preserve first-classness of constraints, 
by which this phenomenon being exhibited as well in the more complicated system of equations upon including a minimally coupled scalar field is rendered less surprising.

Finally, these instances of constraints not being first-class considerations reveal that \cite{HallHaw} is using rather questionable assumptions in its quantization. 
Namely, the Dirac quantization procedure which they use presupposes that each independent subsystem arising from modewise and S, V, T splits has first-class constraints, 
whereas the above argument shows that this does not in fact apply to the S and V pieces.  
Due to this, the Dirac quantization's promotion of classical constraint equations to quantum wave equations 
-- which is a suitable procedure for first-class classical constraint equations -- ceases to be suitable.
Instead, a classical method of dealing with Principles of Dynamics systems possessing aquations in addition to first-class equations is required. 
In particular, a classical reductive treatment prior to quantization 
is suitable as a means of freeing a classical Principles of Dynamics system of equations from containing `specifier equations'.

\section{Details of beables for SIC}

\noindent The {\it Problem of Beables} concerns finding quantities that commute with all the constraints (or maybe just some subset thereof).

Let us start with the pure gravity case.
In Riem$_{0,1,2}$($\mathbb{S}^3$) among the Halliwell--Hawking coordinates themselves, in the modewise case one finds the \K beables $d_{\sn}^{\so}$ and $d_{\sn}^{\se}$. 
The scalar sum (with a different constant of proportionality)
\beq 
s_{\sn}^{\prime} := \mbox{$\frac{1 - \sn^2}{3\sn^2}$}\{a_{\sn} + b_{\sn}\}
\eeq 
is a natural reduced coordinate and this is also a \K beable. 
A further such is 
\beq
\Omega_n = \Omega - A_{\sn}/3 \mbox{ } ,  
\label{Omegan}
\eeq
which also simplifies the blockwise structure of the configuration space metric.

Then note moreover the significant point that the momenta associated with these that are also beables are no longer all momenta conjugate to the purely configurational beables.  
I.e. the two $\pi_{d_{\tn}}$, 
\beq
\pi_{2s\sn} :=  \mbox{$\frac{1}{4}\frac{\sn^2 - 1}{\sn^2 - 4}$}\pi_{b_{\tn}} - \pi_{a_{\tn}} 
\eeq 
and $\pi_{\Omega}$ (specifically a weak \K beable).
Then functionals of these `basis beables' are also classical \K beables.  
The algebra that the basis beables form involves 4 pairs giving brackets of 1 as per the 4-$d$ Heisenberg algebra and the single additional interlinking relation,
\beq 
\mbox{\bf \{}\Omega_{\sn} \mbox{\bf ,} \, \pi_{2s_2} \mbox{\bf \}} = s_{\sn} \mbox{ } . 
\eeq
\noindent For the case with a minimally coupled scalar field, $f_{\sn}$ and 
\beq
\phi^{\prime}_{\sn} := \phi -  3b_{\sn}f_{\sn} \mbox{ } 
\label{phiprimen}
\eeq 
are to be added to this.
The momenta associated with these are then $\pi_{f_{\tn}}$ and the scalar dilational momentum 
\beq
\pi_{2\phi} := \phi\pi_{\phi} + f_{\sn}\pi_{f_{\tn}}
\mbox{ } . 
\eeq 
The further brackets are firstly Heisenberg's 1 for $f_{n}$ and its conjugate, secondly 
\beq
\mbox{\bf \{} \phi_{\sn} \mbox{\bf ,} \, \pi_{2\phi}\mbox{\bf \}} = \phi_{\sn} \mbox{ } ,
\eeq 
and finally the interlinking relations
\beq
\mbox{\bf \{}f_{\sn}     \mbox{\bf ,} \, \pi_{2\phi}   \mbox{\bf \}}       =     f_{\sn}    \mbox{ } , \mbox{ } \mbox{ }  
\mbox{\bf \{}\pi_{2\phi} \mbox{\bf ,} \, \pi_{f_{\tn}} \mbox{\bf \}}       =    \pi_{\sn}  \mbox{ } , \mbox{ } \mbox{ } 
\mbox{\bf \{}\phi_{\sn}  \mbox{\bf ,} \, \pi_{2\sn}    \mbox{\bf \}}       = -  \mbox{$\frac{3}{4}\frac{\sn^2 - 1}{\sn^2 - 4}$}f_{\sn}  \mbox{ } .   
\eeq
\mbox{ } \mbox{ }  It is a furtherly valid point that, given that SIC distinguishes between modewise Superspace on the one hand 
and the result of eliminating the linear constraints on the other hand, \K beables are not the only alternative in choosing a type of beables. 
E.g. in the simpler vacuum case, there is the further issue of beables which (weakly) commute with $\mbox{}^{\sV} \scH$ as well.
Moreover, \K beables remain well-defined, since $\mbox{}^{\sS} \scM_i$ and $\mbox{}^{\sV} \scM_i$ still close as a subalgebra of constraints;\footnote{For beables as objects 
commuting with a set of constraints to make sense, that set of constraints needs to a fortiori close as a subalgebraic structure of the constraints algebraic structure.
Then Dirac beables are always meaningful, but \K beables are only meaningful when the linear constraints in question close among themselves.
See \cite{ABeables} for further discussion and examples.}
the brackets which are problematic are, rather, cross-brackets between some pieces of $\scM_i$ and some pieces of $\scH$.

\section{Spacetime Relationalism} 

\noindent For full GR, the physically irrelevant spacetime transformations are Diff($\FrM$). 
Like Diff($\Sigma$) but contrarily to the Dirac algebroid, these {\sl do} form a Lie algebra:\footnote{Here $\mbox{\bf |[} \, \,  \mbox{\bf ,} \, \, \, \mbox{\bf ]|}$ is the generic 
Lie bracket, Greek letters are spacetime indices and $X^{\mu}, Y^{\mu}$ are smearing functions.}
\be
\mbox{\bf |[} (    \mD_{\mu}    |    X^{\mu})   \mbox{\bf ,} \, (    \mD_{\nu}|Y^{\nu}    ) \mbox{\bf ]|} = (    \mD_{\gamma}    | \, [X, Y]^{\gamma}    ) \mbox{ }  . 
\ee
There is still an issue as to what role $\mD_{\mu}$ plays here; this classical realization of a Lie bracket is {\sl not} conventionally taken to be a Poisson bracket. 
Moreover, unlike with Sec \ref{Intro}'s  Diff($\Sigma$) auxiliaries, these $\mD_{\mu}$ are not conventionally associated with dynamical constraints.      
Nor is the above classical realization of a Lie bracket conventionally taken to be a Poisson bracket. 
Because of that, there is conventionally no complete spacetime analogue of the previous Sec's notion of beables or observables.

\mbox{ }

\noindent It is then  well known that GR perturbation theory can be cast as an unphysical but technically-useful 5-$d$ stack of spacetime 4-geometries that are interrelated via a 
point identification map encoded by the Lie derivative with respect to, now, a {\sl spacetime 4-vector} .

\section{Refoliation Invariance} 

See Fig \ref{Fol-Dep}.

{\begin{figure}[ht]
\centering
\includegraphics[width=0.9\textwidth]{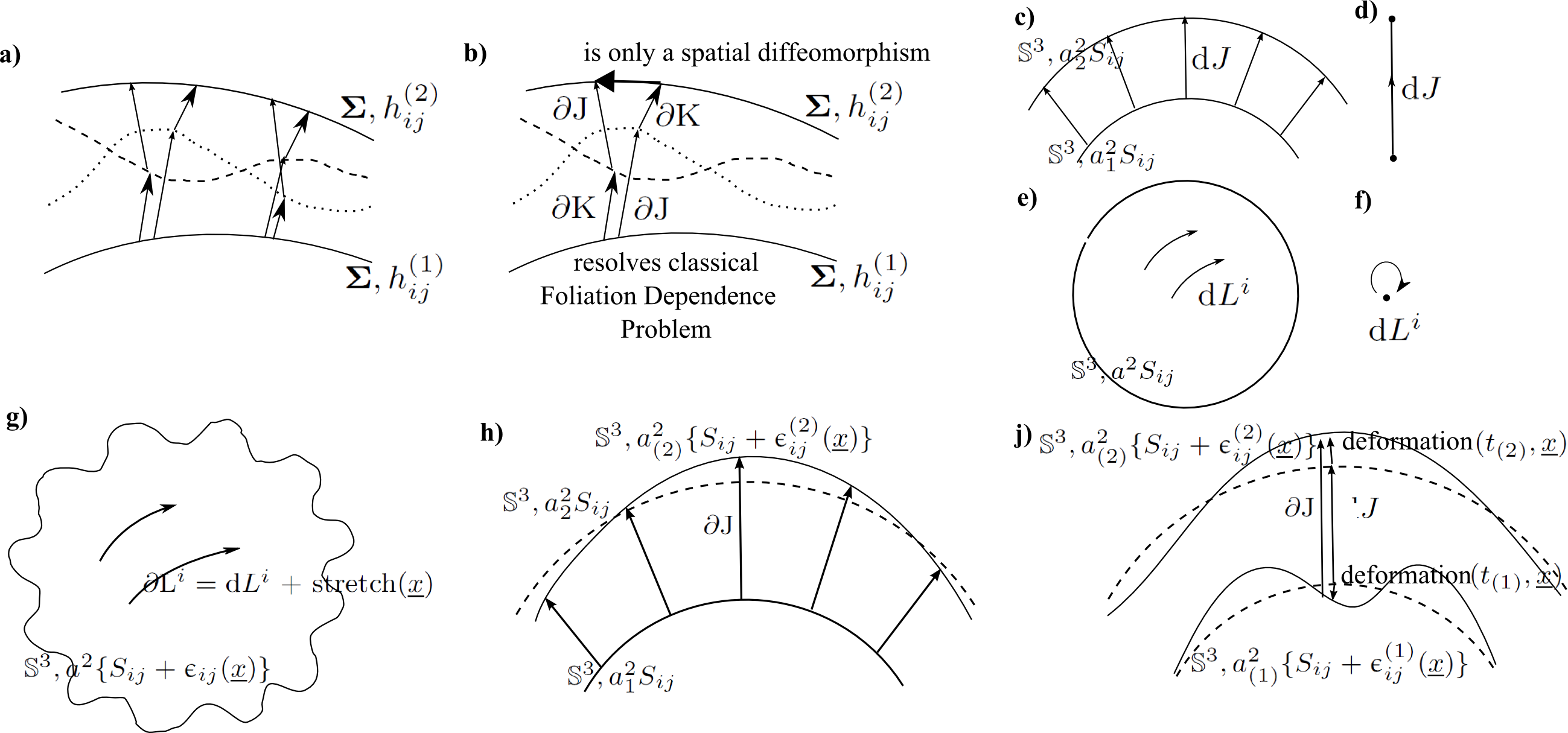}
\caption[Text der im Bilderverzeichnis auftaucht]{\footnotesize{
a) The Foliation Dependence Problem is whether evolution via the dashed spatial hypersurface and via the dotted one give the same physical answers.  
b) Teitelboim's \cite{T73} classical `Refoliation Invariance' resolution of this via the pictorial form of the Dirac algebroid's bracket of two Hamiltonian constraints. 
For minisuperspace \cite{AMSS-1} and mode by mode SIC, this works out trivially.
Although c) and e) over-represent this from a purely minisuperspace perspective compared to d) and f) 
[for which without loss of generality $\d L^i$ = id due to all points being physically identical], c) and e) are needed for comparison with subsequent inhomogeneous perturbations. 
I.e. g) as the slightly bumpy version of e) and h) as some indication of c) in the presence of small deformations. 
In fact the Hamiltonian constraint's action and the evolution are typically between two distinct small deformations of $\mathbb{S}^3$, as indicated in i).  
The reader can then easily imagine the `going via a dotted or dashed choice of a third spatial hypersurface( extension of this progression that nontrivially manifests SIC's 
Foliation Dependence Problem and its classical Refoliation Invariance resolution.}} 
\label{Fol-Dep}\end{figure}} 

\section{Spacetime Construction} 

\noindent {\bf Spacetime Construction Problem} This is here considered in the sense of constructing spacetime from assumptions of just space, 
rather than in the separate sense of arriving at a spacetime continuum from a discrete ontology. 

\mbox{ } 

\noindent At the classical level, space can be embedded into spacetime since the vacuum Einstein field equations are contractions of various well-known embedding equations: 
the Gauss equation, the Codazzi equation and the Ricci equation.
Moreover, constuction of spacetime from space, rather than the a priori assumption of the existence of spacetime, receives substantial further motivation at the quantum level.  
This follows from Quantum Theory implying that fluctuations of configuration are unavoidable. 
However, in the case of GR configurations, these amounts to fluctuations of 3-geometry, and these are too numerous and varied to all be embeddable within a single spacetime.  
Additionally, Wheeler \cite{Battelle, W79} pointed out that the uncertainty principle applies.
Precisely-known position $\underline{q}$ and momentum $\underline{p}$ for a particle are a classical concept corresponding to a worldline.
This perspective breaks down in quantum physics due to the Uncertainly Principle.
Worldlines are replaced by more diffuse wavepackets. 
In the case of GR, then, quantum-level promotions of $\mh_{ij}$ and $\mp^{ij}$ are linked by an uncertainty relation. 
However $\mp^{ij}$ is just a densitized, raised-indices trace-displaced version of the extrinsic curvature $\mK_{ij}$. 
Thus this amounts to $\mh_{ij}$ and $\mK_{ij}$ not concurrently being precisely knowable.
But these are the first and second fundamental forms that form the data for the embedding equations, so this amounts to QM compromising the construction of an embedding into spacetime.  
Thus (something like) the geometrodynamical approach -- in which the set of possible 3-geometries and the dynamics of these is considered -- 
would be expected to take over from spacetime-first approaches at the quantum level.

\mbox{ } 

\noindent Let us for now consider Spacetime Construction at the classical level, 
without assuming that the dynamical features of the geometrodynamics that corresponds to GR are a priori known.
There is then the following `spacetime from space' result for full GR \cite{RWR, AM13}.
Let $\uppsi^A$ denote fundamental-field second-order minimally-coupled bosonic matter, with conjugate momenta $\Pi^{\uppsi}_{A}$.  
Then $\pa \ms^{\sg\sr\sa\sv-\uppsi} = \sqrt{\pa \ms^{\sg\sr\sa\sv \, 2}_{y, w} + \sum_{\uppsi}y^{-1}_{\uppsi}\pa \ms^2_{\uppsi}}$
(minimal coupling gives no metric--matter kinetic cross-terms so it decomposes in this blockwise manner).
Also here, $\pa \ms_{\uppsi}^2 := M_{AB}\pa \uppsi^A\pa \uppsi^B$ 
for matter configuration space metric $M_{AB}$ that I take to be ultralocal in the metric and with no dependence on the matter fields themselves.
Additionally, $\mW^{\sg\sr\sa\sv-\uppsi} := a \, \mR + b + \sum_{\uppsi}a_{\uppsi} \mU_{\uppsi}$ for $\mU_{\uppsi}$ minus the matter sector's potential.  
This can only depend on the spatial derivatives of the spatial metric through the spatial Christoffel symbols.  
It is logical to give the expressions for the conjugate momenta here, rather than just symbol-defining them below
Then the generalization of the GR Hamiltonian constraint is 
\beq
\scH_{x, y, y_{\uppsi}, a, a_{\uppsi}, b} := \{y\{\mh^{ik}\mh^{jl} - x \mh^{ij}\mh^{kl}/2\}\mp_{ij}\uppi_{kl} 
+ \sum\mbox{}_{\mbox{}_{\mbox{\scriptsize $\uppsi$}}}  y_{\uppsi}  M^{AB}  \Pi_{A}\Pi_{B}\}/\sqrt{\mh} - \sqrt{\mh}\left\{ a\,\mR + b 
+ \sum\mbox{}_{\mbox{}_{\mbox{\scriptsize $\uppsi$}}}  a_{\uppsi} \mU_{\uppsi}\right\} = 0 \mbox{ } , 
\eeq 
where $\mp$ is the trace of $\mp^{ij}$. 
For these models, changes in all the matter degrees of freedom do have the opportunity to contribute to the emergent time standard, 
$t^{\se\sm(\sJ\sB\sB)} = \int \pa s^{\sg\sr\sa\sv-\uppsi}/\sqrt{2 \mW^{\sg\sr\sa\sv-\uppsi}}$.  
Then computing the bracket of $\scH_{x, y, y_{\uppsi}, a, a_{\uppsi}, b}$ with itself gives -- after a quite lengthy calculation along the lines of \cite{AM13} --
$$
\mbox{\bf \{} 
(    \scH_{x, y, y_{\uppsi}, a, a_{\uppsi}, b}    |    \pa \mJ    ) 
\mbox{\bf ,} \, 
(    \scH_{x, y, y_{\uppsi}, a, a_{\uppsi}, b}    |    \pa \mK    )
\mbox{\bf \}}
=  
\left(
ay \{  \scM_i^{\sg\sr\sa\sv-\uppsi} + 2 \{1 - x\} \mD_i \uppi \}
+ \sum_{\uppsi}
\left\{ 
\underline{ ay
\left\lfloor
\Pi^{\sfA}\frac{\delta\pounds_{\pa{\underline{\sL}}} \uppsi_{\sfA}}{\delta\pa{\mL}^i}
\right\rfloor } 
\right.
\right.
$$
\beq
\mbox{ } \mbox{ } \mbox{ } \mbox{ } \mbox{ } \mbox{ } \mbox{ } \mbox{ } \mbox{ } 
\mbox{ } \mbox{ } 
\left.
\left.
       \left.
- 2 a_{\uppsi}y_{\uppsi} \underline{        M^{AB} \Pi_{A}  \frac{    \pa \mU_{\uppsi}    }{   \pa \, \pa_i\uppsi^{B}   }        }    
      \right\}
	   - 2\underline{y        
			\left\{
 \mp_{jk} - \frac{x}{2}\mp \mh_{jk}
            \right\}
\mh_{il}
\sum_{\uppsi}a_{\uppsi}            
			\left\{
            \frac{    \pa \mU_{\uppsi}    }{    \pa {\Gamma^c}\mbox{}_{jl}    }\mh^{ck} - \frac{1}{2}\frac{    \pa \mU_{\uppsi}    }{    \pa{\Gamma^c}\mbox{}_{jk}    }\mh^{lc}
            \right\}    }
\right| 
\pa \mJ \, \overleftrightarrow{\pa}^i  \pa \mK 
\right)
\mbox{ } .
\label{4-Factors-Matter}
\eeq
Let us first explain that the `floor bracket' $\lfloor \mbox{ } \rfloor$ is used to denotes the extent to which the variational derivative inside of this bracket acts. 
Since the intent of this calculation is to cover conventionally-used fundamental matter fields such as minimally-coupled scalars, Electromagnetism etc, 
I point out that all of these have no Christoffel symbol terms in their potentials, by which the last underlined grouping drops out.

The pure-gravitation parts of these results then follow from the second term \cite{RWR, AM13}. 
This is an obstruction term with multiple factors, each of which being zero provides a distinct option.  
The local relativity of matter parts of these results (\cite{AM13} and references therein) follow as means of the other two underlined terms cancelling each other.  
This result can be interpreted as a restriction -- via a Dirac-type procedure \cite{Dirac, AM13} -- for what type of consistent theories can emerge within the ansatz made.  
One of these options then fixes the inverse DeWitt supermetric value $x = 1$ alongside the locally Lorentzian physics of SR emerging. 
The $y$'s being zero furnishes a distinct option another option -- a geometrostatics that is in a sense a Riemannian space generalization of locally Galilean relativity.
The $a$'s being zero give rise to a third option, which gives strong-coupled limit of GR alongside Carrolian relativity (here each point can only communicate with itself). 
These second and third options are interesting through their possessing alternative local relativities, but are clearly not realized by the world around us.  
Finally $\uppi/\sqrt{h} = constant$ arises as a fourth option -- constant mean curvature (CMC) slicing -- which is not furtherly pursued here.
The $x = 1$ option furthermore leads to a recovery of the GR-type notion of spacetime.
In essence, if this option is followed, then the constraint equations admit interpretation as embedding equations into a manifold of dimension one larger.

\mbox{ } 

\noindent In minisuperspace and modewise SIC, the above result becomes trivial, 
due to the $\mD_i$ contained in a cofactor of $x - 1$ having nothing to act upon.  
Due to this, in the minisuperspace version of the working, the a priori free coefficient $x$ in the supermetric does not get fixed to 1.
Thus minisuperspace is not a sufficiently complex model arena in which to investigate Spacetime Construction; 
it also possesses surfaces privileged by homogeneity which `pile up' in a simpler manner than generic GR's geometrodynamical strutting structure \cite{MTW}.

The current Article's new result is, rather the specific case of (\ref{4-Factors-Matter}) for the SIC model at the level of plain 
(rather than furthermore modewise and S, V, T split) perturbation theory.  
In this setting, (\ref{4-Factors-Matter}) survives in the following nontrivial form for the bracket of $\scH_{a,b,x,y}$ with itself:  
$$
\mbox{\bf \{} 
(      \scH_{x, y, y_{\upphi}, a, a_{\upphi}, b}    |    \pa \mJ    ) 
\mbox{\bf ,} \, 
(      \scH_{x, y, y_{\upphi}, a, a_{\upphi}, b}    |    \pa \mK    ) 
\mbox{\bf \}}_{2}
=  
\left(
ay \{   \scM_{1i}^{\sg\sr\sa\sv-\upphi}    + 2    \{1 - x\}    \mbox{exp}(-2\Omega)      \mD_i\uppi_1  \}
+
ay
\left\lfloor
\pi_1^{\upphi}   \frac{\delta\pounds_{\d{\underline{\sL}}}\upphi_0}{\delta\d{\mL}^i}
\right\rfloor 
\right.
$$
\beq
\mbox{ } \mbox{ } \mbox{ } \mbox{ } \mbox{ } \mbox{ } \mbox{ } \mbox{ } \mbox{ } 
\mbox{ } \mbox{ } \hspace{2.5in}
\left.
\left.
\left.
- 2 a_{\upphi} y_{\upphi} \mbox{exp}(-3\Omega) \pi^{\upphi}_0 \frac{\pa \mU_{1\,\upphi}}{\pa\,\pa_i\upphi}    
\right\}
\right| S^{ij} 
\{ \pa\mK_0 \, \pa_j \, \pa\mJ_1  - \pa\mJ_0 \, \pa_j \, \pa\mK_1 \}  
\right)
\mbox{ } .
\eeq
By this result, it is clear that the SIC model arena presents a fully functional form for classical Spacetime Construction from the assumption of spatial structure. 
Indeed, all four of the above-mentioned options -- GR with local Lorentzian relativity, geometrostatics with Galilean relativity, strong gravity with Carrollian relativity, and 
CMC sliced theories, are already represented within this SIC model arena.  
Due to this, this model arena is of substantial further use in investigating Spacetime Construction the assumption of spatial structure.

\section{Conclusion} 

This Article concerned a local resolution of the Problem of Time (PoT). 
`A local' means that the Global Problems of Time and Multiple Choice Problems of Time remain.
On the other hand, Temporal, Configurational and Spacetime Relationalisms, Constraint Closure, Beables, Foliation Dependence and Spacetime Construction have been dealt with at the 
classical level.
Whilst \K commented that \cite{Kuchar99} the Halliwell--Hawking model would exhibit `{\it `a small flexibility to wiggle the instants}" of the foliation, 
he did not provide any detailed results for the PoT in such models. 
This Article supplies some of those, and in doing so delineates how Refoliation Invariance and Spacetime Construction nontriviality require the plain (rather than modewise) analysis.  
These two differ because mode expansions are of the spheres of homogeneity themselves.  
Thus, whereas one can refoliate FLRW spacetime in whatever other manner (paralleling Dirac's insight for Minkowski spacetime treated thus \cite{Dirac}), 
one cannot expect the modewise split to carry over to other foliations.
Hence use of the modewise split amounts to an extra use of privileged structure.

In the slightly inhomogeneous cosmology arena, at the classical level the modewise approach leads to a timefunction that bears small corrections relative to the usual concept of cosmic time.
These are Machian corrections due to their taking into account the effects of small inhomogeneities. 
This approach additionally involves passing to a reduced formulation.
This reduced formulation [in particular (\ref{An}, \ref{Omegan}, \ref{phiprimen}) arising] can be viewed as a classical precursor of Wada's approach \cite{Wada} to quantization \cite{HallHaw}, 
as opposed to \cite{HallHaw}'s own unreduced approach; see \cite{ABook} for more details of these inter-relations.
Also note that the modewise constraints behave in a perhaps unexpected manner as compared to the full GR constraints; in particular, there are two second-class pairs.  
I further provide this slightly inhomogeneous cosmology model with a set of basis beables out of which \K beables can be constructed.

The Machian classical Frozen Formalism Problem resolution is broken at the quantum level \cite{A13}. 
However, its methodology can be started afresh at this level \cite{ACos2}. 
Furthermore, the difference between the two timestandards thus produced can itself be seen to be Machian (time is now emergent from {\sl quantum} change).  
I provided this at the semiclassical quantum level for RPM's and minisuperspace models in \cite{ACos2, QuadII, AMSS-1} as preparation for doing so for slightly inhomogeneous cosmology 
in the present and forthcoming articles. 
Moreover, semiclassical approaches to the PoT rely on the WKB approximation, which requires its own further justification. 
One approach to this is that decohereing histories \cite{HartleIL} provides such a regime.
Histories, the semiclassical approach and timeless records inter-protect each other to a greater extent as a three-way combined scheme \cite{H03, H09H11, AHall, A13, FileR}.

Some further interesting questions are as follows. 

\noindent 1) The following factorization of the strategizing occurs at the classical level. 
Dirac's algebroid addresses all of Constraint Closure, Foliation Dependence and Spacetime Construction. 
On the other hand, resolving Best Matching gives also resolutions to Temporal Relationalism and \K beables.
Then which of this article's arguments and results about brackets carry over to the quantum level?
And does such a factorization continue to apply in the case of slightly inhomogeneous cosmology at the semiclassical level?  

\noindent 2) Which subalgebra of functionals of the basis of \K beables can be consistently promoted to quantum \K beables?

\noindent 3) A limitation of the current Article's mode by mode slightly inhomogeneous cosmology arena is that its perturbative split carries background split problems. 
These are not exactly the same as those documented for background Minkowski spacetime splits in quantum gravity, but some are similar.  
Nonperturbative midisuperspace models are better in this respect.  

\noindent 4) Another future direction is to push the present program to third-order workings. 
S, V, T decoupling ends here.
Conventional cosmological calculations have been extended to this qualitatively more general case, albeit without semiclassical time emerging from its calculations.

\noindent 5) Global Problems of Time \cite{Kuchar92I93, APoT3}  affect multiple facets and multiple of the strategies considered in this Article.
E.g. difficulties with choosing an 'everywhere-valid' timefunction, linear constraint resolution, beables, Refoliation Invariance resolution, Spacetime Construction...

\noindent 6) Multiple Choice Problems \cite{Kuchar92I93}.  
Canonical equivalence of classical formulations of a theory \cite{Gotay} does not imply unitary equivalence of the quantizations of each.    
By this, different choices of timefunction can lead to inequivalent quantum theories. 
This also applies as regards choices of beables.  

\mbox{ }

\noindent {\bf Acknowledgements} 
I thank Julian Barbour, Jonathan Halliwell, Marc Lachi$\grave{\me}$ze-Rey and Flavio Mercati for discussions and 
John Barrow, Jeremy Butterfield, Malcolm MacCallum, Don Page and Reza Tavakol for helping me with my career.


\end{document}